\documentclass[5p,authoryear,times]{elsarticle}

\usepackage{amsmath}
\usepackage{amssymb}
\usepackage{txfonts}
\usepackage{epsfig,graphicx}
\usepackage{verbatim}
\usepackage{bm}
\usepackage{url}

\usepackage[colorlinks=true, linkcolor=blue, citecolor=blue, urlcolor=blue, draft]{hyperref}

\bibpunct{(}{)}{;}{a}{}{,} 

\journal{Astronomy and Computing}

\DeclareTextFontCommand{\mytexttt}{\ttfamily\hyphenchar\font=45\relax}

\def\tablefoot#1{\par\vspace*{2ex}%
 \parbox{\hsize}{\leftskip0pt\rightskip0pt
 {\noindent\small\textbf{Notes.}~#1\par}}}

\makeatletter
\def\@author#1{\g@addto@macro\elsauthors{\normalsize%
    \def\baselinestretch{1}%
    \upshape\authorsep#1\unskip\textsuperscript{%
      \ifx\@fnmark\@empty\else\unskip\sep\@fnmark\let\sep=,\fi
      \ifx\@corref\@empty\else\unskip\sep\@corref\let\sep=,\fi
      }%
    \def\authorsep{\unskip,\space}%
    \global\let\@fnmark\@empty
    \global\let\@corref\@empty  
    \global\let\sep\@empty}%
    \@eadauthor={#1}
}
\makeatother

\begin{document}

\begin{frontmatter}

\title{cuFFS: A GPU-accelerated code for Fast Faraday Rotation Measure Synthesis.}

\author[kap,ast,csi]{S.S. Sridhar\corref{cor1}}
\cortext[cor1]{Corresponding author. E-mail address: sarrvesh.ss@gmail.com}
\ead{sarrvesh.ss@gmail.com}
\author[csi]{G.~Heald}
\author[kap]{J.M.~van der Hulst}

\address[kap]{Kapteyn Astronomical Institute, University of Groningen, Postbus 800, 9700AV Groningen, The Netherlands.}
\address[ast]{ASTRON, the Netherlands Institute for Radio Astronomy, Postbus 2, 7990 AA, Dwingeloo, The Netherlands.}
\address[csi]{CSIRO Astronomy and Space Science, 26 Dick Perry Ave, Kensington, WA 6151, Australia.}

\begin{abstract}
Rotation measure (RM) synthesis is a widely used polarization processing algorithm for reconstructing polarized structures along the line of sight. Performing RM synthesis on large datasets produced by telescopes like LOFAR can be computationally intensive as the computational cost is proportional to the product of the number of input frequency channels, the number of output Faraday depth values to be evaluated and the number of lines of sight present in the data cube. The required computational cost is likely to get worse due to the planned large area sky surveys with telescopes like the Low Frequency Array (LOFAR), the Murchison Widefield Array (MWA), and eventually the Square Kilometre Array (SKA). The massively parallel General Purpose Graphical Processing Units (GPGPUs) can be used to execute some of the computationally intensive astronomical image processing algorithms including RM synthesis. In this paper, we present a GPU-accelerated code, called cuFFS or CUDA-accelerated Fast Faraday Synthesis, to perform Faraday rotation measure synthesis. Compared to a fast single-threaded and vectorized CPU implementation, depending on the structure and format of the data cubes, our code achieves an increase in speed of up to two orders of magnitude. During testing, we noticed that the disk I/O when using the Flexible Image Transport System (FITS) data format is a major bottleneck and to reduce the time spent on disk I/O, our code supports the faster HDFITS format in addition to the standard FITS format. The code is written in C with GPU-acceleration achieved using Nvidia's CUDA parallel computing platform. The code is available at \url{https://github.com/sarrvesh/cuFFS}.

\end{abstract}


\end{frontmatter}

\section{Introduction}
Until the early 2000s, improvements in the execution speeds of scientific programs were largely due to the increase in CPU clock rates. This however changed in last decade when chip manufacturers realized that it was hard to keep increasing the clock rate as it led to overheating of the chips due to thermal losses \citep[see for example][]{patterson2014}. 

Processor manufacturers work around the ``thermal wall'' by implementing multiple processing units (or cores) on a single chip leading to the design of multicore processors. While each core on the chip is not faster than its predecessor, multicore processors achieve higher compute throughput using parallelization. GPUs extend this strategy of implementing multiple cores on a single chip to an extreme limit. In comparison to multicore CPUs, GPUs achieve high compute throughput by maximizing the number of processing units (or cores) on a chip. For example, NVIDIA's Tesla P100 card has 56 streaming microprocessors each with 64 processing units resulting in a total of 3584 single precision (or 1792 double precision) floating point compute cores on a single chip. An added advantage of using GPUs over CPU compute clusters is that GPUs consume less energy measured in Watts per FLOP\footnote{Floating point operation} \citep{trancoso2005,michalakes2008}.

While GPUs were originally designed to be used as graphics coprocessors\footnote{Coprocessors are secondary processors mounted on a device that complements the compute capabilities of the primary processor.} to handle higher resolution and display rates demanded by the gaming industry, the massive compute capability of GPUs has been exploited leading to the creation of a new field called GPGPU (General Purpose GPU) computing \citep{owens2007}. While the initial scientific programs that were run on GPUs had to be ported to the GPU native programming languages, software solutions made available in the last decade have made it far easier to develop GPU codes \citep{cuda}. A review of GPU use in scientific computing can be found in \cite{owens2007} and \cite{owens2008}. 

Graphical processing units have been used quite extensively in astronomy in areas of research ranging from real-time applications like adaptive optics \citep[for example, see][]{bouchez2012} and correlators for radio interferometry \citep{schaaf2004}, easily parallelizable algorithms like N-body simulations \citep{zwart2007,elsen2007}, to more complex algorithms in general relativistic magnetohydrodynamics \citep{zink2011}. For a detailed discussion on the use of GPUs in astronomy, see \cite{fluke2011} and \cite{fluke2012}.

Within the field of radio astronomy, GPUs have largely been restricted to implementing fast real-time signal processing applications like correlators and real-time calibration pipelines\citep[for example, see][]{schaaf2004, harris2008, ord2009, chennamangalam2014, price2016}, and pulsar de-dispersion pipelines \citep{magro2011}. Recently, the Dynamic Radio Astronomy of Galactic Neutron stars and Extragalactic Transients \citep[DRAGNET;][]{bassa2017}\footnote{\url{www.astron.nl/dragnet/}} project with the LOFAR radio telescope has started using GPUs to search for fast radio transients in real-time. GPUs have also been used to accelerate solving the Radio Interferometer Measurement Equation \citep{hamaker1996} to calibrate radio interometry data \citep{perkins2015}. Within the context of medical imaging, it has been demonstrated that the gridding procedure that needs to be carried out before applying two dimensional inverse Fourier transform to the data can be accelerated using GPUs \citep{schomberg1995,schiwietz2006}. Furthermore, a GPU implementation of the fast fourier transform library cuFFT is available as part of the official CUDA toolkit.

In this paper, we present a GPU-accelerated program -- CUDA-accelerated Fast Faraday Synthesis (cuFFS) -- to perform a commonly used technique in radio polarimetry called Faraday rotation measure (RM) synthesis. This paper is organized as follows: The theoretical background and computational complexity of the rotation measure synthesis method are presented in section \ref{sec:background}. Our GPU implementation of the RM Synthesis code is explained in section \ref{sec:gpuimplementation} and benchmark tests are presented in section \ref{sec:benchmarks}. Finally, we present our conclusion and future prospects for improvements to the software package in \S \ref{sec:conclusion}.

\section{Background} \label{sec:background}
\subsection{RM synthesis in theory} \label{sec:rmsynthesis}
Linearly polarized synchrotron emission is an important observational probe to study astrophysical magnetic fields. Synchrotron emission is produced by cosmic ray electrons accelerating across magnetic field lines. The polarization angle of the synchrotron radiation depends on the orientation of the magnetic field in the plane of the sky. Thus resolved polarimetric observations allow us to probe and study the structure of astrophysical magnetic fields\footnote{Polarized synchrotron emission probes only the ordered component of the magnetic field. Distinction between different components of astrophysical magnetic fields and the observational probes used to study them is beyond the scope of this paper. For a recent review on this topic, see \cite{fletcher2015} and \cite{beck2016}.}. However, in practice this is complicated owing to the fact that the observed polarization angle and polarization fraction\footnote{Polarization fraction is defined as polarized intensity divided by total intensity of the synchrotron radiation.} is different from the intrinsic properties at the source due to propagation effects caused by Faraday rotation.

Faraday rotation is a phenomenon by which the plane of polarization of a linearly polarized electromagnetic wave gets rotated due to circular birefringence as it propagates through a magneto-ionic medium like the interstellar medium or the ionosphere. The observed polarization angle $(\chi)$ is related to the intrinsic polarization angle $(\chi_0)$ through a quantity called Faraday rotation measure $(RM)$ as
\begin{equation}\label{eq:PAdefn}
\chi(\lambda^2) = \chi_0 + RM \lambda^2
\end{equation}
where $RM$ is related to the strength of the magnetic field projected along the line of sight $(B_{||} = \vec{B} \cdot \mathrm{d}\vec{l})$ as
\begin{equation}
RM = 0.81 \int^{observer}_{source} n_e \vec{B} \cdot \mathrm{d}\vec{l}.
\end{equation}
In the above equation, $n_e$ is the number density of electrons measured in units of cm$^{-3}$, $l$ is the pathlength in parsec, $B$ is in $\mu$G, and $\phi$ is in rad m$^{-2}$.

The observed complex polarization vector $P(\lambda^2)=Q(\lambda^2) + iU(\lambda^2)$ is related to the observed polarization angle $\chi(\lambda^2)$, fractional polarization $p$, and total intensity $I$ as
\begin{equation}\label{eq:Pdefn}
P(\lambda^2) = pIe^{2i\chi(\lambda^2)}.
\end{equation}
Note that equations \ref{eq:PAdefn} and \ref{eq:Pdefn} allow us to relate the observed polarization vector with the intrinsic polarization angle and the Faraday rotation measure. Replacing the Faraday rotation measure with a generalized Faraday depth $\phi$, and since emission from multiple $\phi$ can contribute to the observed polarization vector $P(\lambda^2)$, following \cite{burn1966} we can write
\begin{equation} \label{eq:Predefn}
P(\lambda^2) = \int^{+\infty}_{-\infty} pIe^{2i[\chi_0 + \phi\lambda^2]} \mathrm{d}\phi = \int^{+\infty}_{-\infty} F(\phi) e^{2i\phi\lambda^2} \mathrm{d}\phi.
\end{equation}
where $F(\phi)$ is defined as the Faraday Dispersion Function $F(\phi) := pIe^{2i\chi_0}$. Inverting this Fourier transform-like equation, 
\begin{equation} \label{eq:Fdefn}
F(\phi) = \int^{+\infty}_{-\infty} P(\lambda^2)e^{-2i\phi\lambda^2} \mathrm{d}\lambda^2
\end{equation}
we can see that the Faraday Dispersion Function (FDF) represents the emitted polarized flux as a function of Faraday depth. Faraday depth is related to the strength of the magnetic field lines projected along the line of sight through the relation

\begin{equation}
\phi(r) = \int^{observer}_r n_e \vec{B} \cdot \mathrm{d} \vec{l}
\end{equation}
and is equal to $RM$ only for the special case when there is a single Faraday rotating medium with no internal Faraday rotation between the source and the observer \citep{vallee1980, brentjens2005}. From equations \ref{eq:Predefn} and \ref{eq:Fdefn}, it is easy to see that $P(\lambda^2)$ and $F(\phi)$ form a Fourier-like transform pair. 

Recovering $F(\phi)$ from the observed $P(\lambda^2)$ is however not straight-forward as we cannot measure $P(\lambda^2)$ for $\lambda^2<0$. Furthermore, $P(\lambda^2)$ is also not measured for all possible values of $\lambda^2>0$. Rotation measure synthesis is a technique that was formulated by \cite{brentjens2005} which tries to recover $F(\phi)$ using a discrete sampling of $P(\lambda^2)$. In addition to reconstructing $F(\phi)$, the procedure also improves the sensitivity of the polarization measurement by adding up derotated polarization signal across the entire observed bandwidth. In the following section, we provide a brief overview of the procedure. For a detailed account on RM synthesis, we refer the reader to \cite{brentjens2005} and \cite{heald2009}.

The rotation measure synthesis technique works around the problem of incomplete $\lambda^2$ coverage by introducing a window function $W(\lambda^2)$ which is non-zero only for $\lambda^2$ values at which complex polarization $P(\lambda^2)$ has been measured. Thus, equation \ref{eq:Predefn} can be modified as
\begin{equation}
\widetilde{P}(\lambda^2) = W(\lambda^2) P(\lambda^2) = W(\lambda^2) \int^{+\infty}_{-\infty} P(\lambda^2)e^{-2i\phi\lambda^2} \mathrm{d}\lambda^2
\end{equation}
Combining the above equation with equation \ref{eq:Fdefn} and applying the convolution theorem, the reconstructed Faraday dispersion function $\widetilde{F}(\phi)$ is related to the ``true'' Faraday dispersion function as
\begin{equation} \label{eq:fPhi}
\widetilde{F}(\phi) = F(\phi) * R(\phi) = K \int^{+\infty}_{-\infty} \widetilde{P}(\lambda^2) e^{-2i\phi \lambda^2} \mathrm{d}\lambda^2
\end{equation}
where
\begin{equation} \label{eq:kPhi}
K = \left( \int^{+\infty}_{-\infty} W(\lambda^2) \mathrm{d}\lambda^2 \right)^{-1}.
\end{equation}
From the above equation, it is easy to see that the recovered Faraday dispersion function $(F(\phi))$ is a convolution of the ``true'' Faraday dispersion function with smoothing kernel $(R(\phi))$ called the Rotation Measure Spread Function (RMSF). Conceptually, this is analogous to synthesis imaging where $\widetilde{F}(\phi)$ is equivalent to the dirty image and $R(\phi)$ is equivalent to the dirty beam. The RMSF is related to the weight function $W(\lambda^2)$ as
\begin{equation} \label{eq:rPhi}
R(\phi) = K \int^{+\infty}_{-\infty} W(\lambda^2) e^{-2i\phi \lambda^2} \mathrm{d}\lambda^2.
\end{equation}

\subsection{Practical implementation of RM synthesis} \label{sec:computationcosts}
\begin{figure*}
\centering
\resizebox{\hsize}{!}{\includegraphics{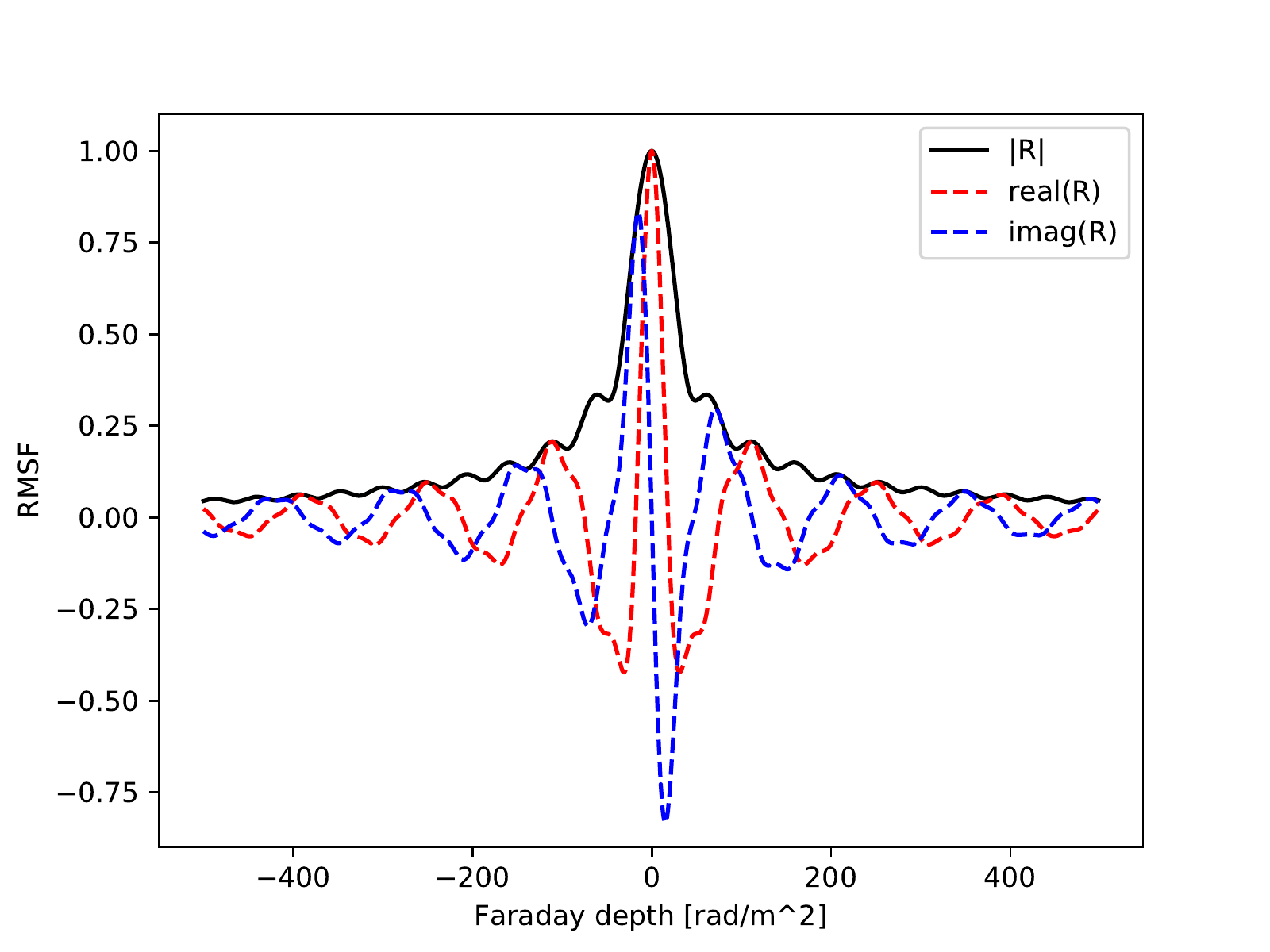}\includegraphics{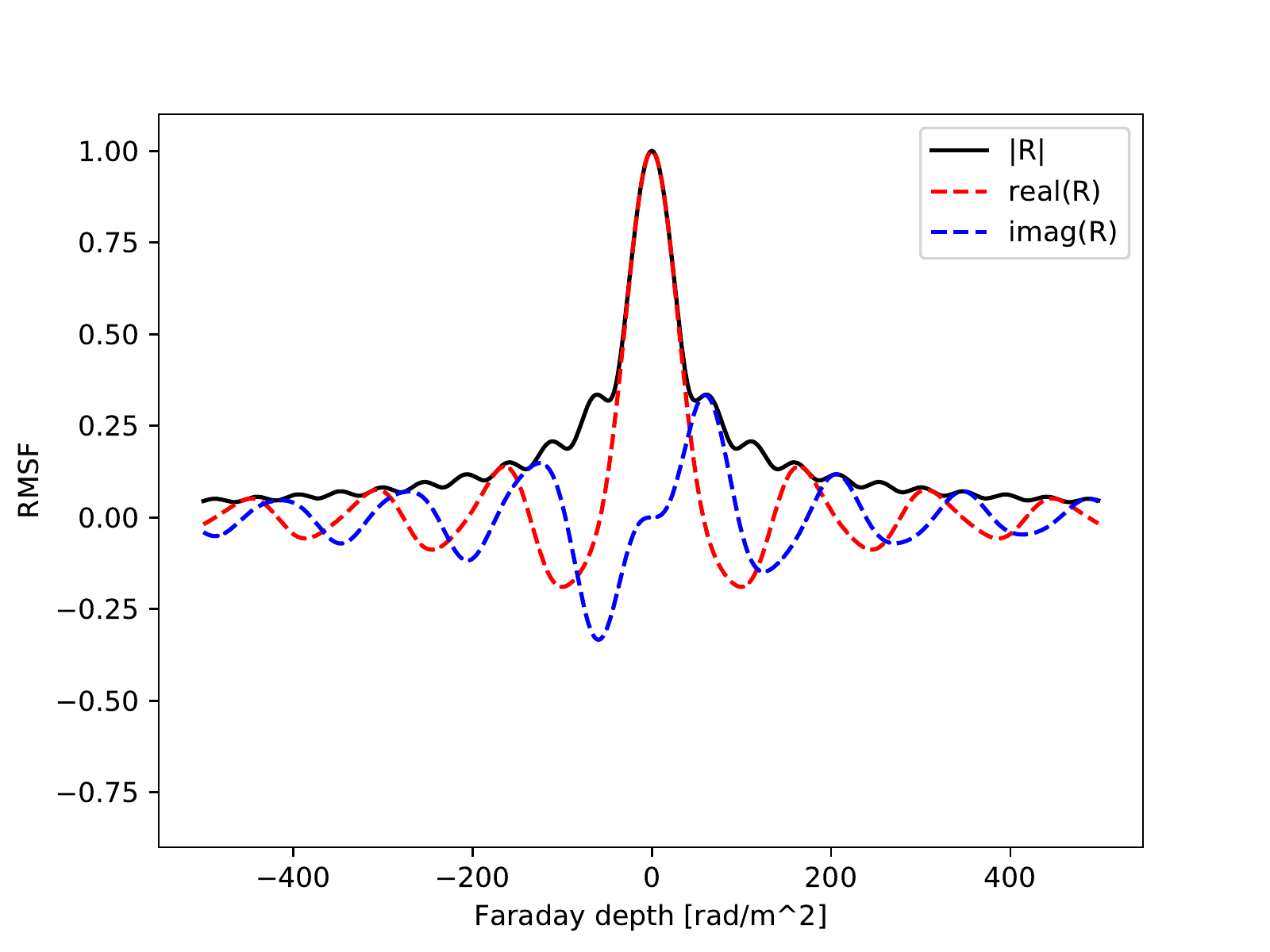}}
\caption{Rotation Measure Spread Function (RMSF) for a mock dataset consisting of 1000 frequency channels covering the 1 -- 2 GHz frequency range with $\lambda_0^2 = 0$ (\textit{left}) and $\lambda_0^2 \neq 0$ (\textit{right}). Notice that introducing the $\lambda_0^2$ term reduces the rapid rotation of $\widetilde{Q}(\phi)$ and $\widetilde{U}(\phi)$.}
\label{fig:rmsf}
\end{figure*}

As discussed in the previous section, the complex polarization vector is not measured for all values of $\lambda^2$. Modern radio telescopes measure $P(\lambda^2)$ in narrow frequency channels spread across a wide bandwidth each with equal channelwidth, expressed in $\lambda^2$ space as $\delta \lambda^2$. In such a scenario, the integrals in equations \ref{eq:fPhi}, \ref{eq:kPhi} and \ref{eq:rPhi} can be replaced by sums over each frequency channel provided $\phi \delta \lambda^2 \ll 1$:
\begin{equation} \label{eq:Fsum}
\widetilde{F}(\phi) \approx K \sum^{N}_{i=1} \widetilde{P}_i e^{-2i\phi(\lambda^2_i - \lambda^2_0)}
\end{equation}
\begin{equation} \label{eq:Rsum}
R(\phi) \approx K \sum^{N}_{i=1} w_i e^{-2i\phi(\lambda^2_i - \lambda_0^2)}
\end{equation}
\begin{equation} \label{eq:Ksum}
K = \left( \sum^{N}_{i=1} w_i \right)^{-1}
\end{equation}
In the above equations, $\widetilde{P}_i$ is the complex polarization vector measured at $\lambda_i$ and $w_i = W(\lambda_i)$. Figure \ref{fig:rmsf} shows a RMSF calculated using equation \ref{eq:Rsum} for a mock dataset consisting of 1000 frequency channels covering the 1 -- 2 GHz frequency range (L-band). Notice in equations \ref{eq:Fsum} and \ref{eq:Rsum}, we have introduced an additional term $\lambda_0^2$ in the exponent which is set to the mean of $\lambda_i^2$. Introducing this additional $\lambda_0^2$ term is useful as a cosmetic addition as it prevents rapid rotation of $\widetilde{Q}(\phi)$ and $\widetilde{U}(\phi)$ as can be seen in Figure \ref{fig:rmsf}. This cosmetic addition allows one to make an accurate estimate of $P_\mathrm{peak}$ and thus make an accurate estimate of the polarization angle $\chi$ using the relation

\begin{equation}
\chi = 0.5 \tan^{-1}\left( \frac{U_{\phi\mathrm{peak}}}{Q_{\phi\mathrm{peak}}} \right) e^{-2i\phi_\mathrm{peak}\lambda^2_0}.
\end{equation} 
Note that $Q_{\phi\mathrm{peak}}$ and $U_{\phi\mathrm{peak}}$ are the values of $\widetilde{Q}(\phi)$ and $\widetilde{U}(\phi)$ at Faraday depth $\phi_\mathrm{peak}$ where $|\widetilde{F}(\phi)|$ is maximum.

The aim of the RM synthesis software package is to compute equations \ref{eq:Fsum} and \ref{eq:Ksum}. Equation \ref{eq:Rsum} is needed only if one wishes to deconvolve $R(\phi)$ from the reconstructed Faraday dispersion function using a procedure called RM-CLEAN\footnote{RM-CLEAN is not supported in the current version of cuFFS. We intend to implement RM-CLEAN and other advanced polarization processing techniques in the future.} \citep{heald2009}. 

Since $\widetilde{F}(\phi)$ and $\widetilde{P}(\lambda^2)$ in equation \ref{eq:Fsum} are complex vectors, the equation can be written in terms of its Stokes Q and Stokes U components as
\begin{equation} \label{eq:qPhi}
\widetilde{Q}(\phi) = K \sum^{N}_{i=1} Q_{\lambda i} \cos2\phi(\lambda_i^2 - \lambda_0^2) + U_{\lambda i} \sin2\phi(\lambda_i^2 - \lambda_0^2)
\end{equation}
\begin{equation} \label{eq:uPhi}
\widetilde{U}(\phi) = K \sum^{N}_{i=1} U_{\lambda i} \cos2\phi(\lambda_i^2 - \lambda_0^2) - Q_{\lambda i} \sin2\phi(\lambda_i^2 - \lambda_0^2).
\end{equation}
\begin{equation} \label{eq:pPhi}
\widetilde{F}(\phi) = \sqrt{\widetilde{Q}^2(\phi) + \widetilde{U}^2(\phi)}
\end{equation}

The above two equations compute the recovered Faraday dispersion function for a given value of $\phi$. In practice, the above equations need to be computed for a wide range of values of $\phi$ to get a sense for the variation of the intrinsic polarization vector as a function of Faraday depth. Thus the above equations become
\begin{multline}
\widetilde{Q}(\phi_j) = K \sum^{N}_{i=1} Q_{\lambda i} \cos2\phi_j(\lambda_i^2 - \lambda_0^2) + U_{\lambda i} \sin2\phi_j(\lambda_i^2 - \lambda_0^2); \\ \forall \phi_j \in [\phi_\mathrm{min}, \phi_\mathrm{max}] 
\end{multline}
\begin{multline}
\widetilde{U}(\phi_j) = K \sum^{N}_{i=1} U_{\lambda i} \cos2\phi_j(\lambda_i^2 - \lambda_0^2) - Q_{\lambda i} \sin2\phi_j(\lambda_i^2 - \lambda_0^2); \\ \forall \phi_j \in [\phi_\mathrm{min}, \phi_\mathrm{max}].
\end{multline}
These two equations need to be computed for each line of sight in a radio image to construct a 3D RM-cube that contains the Faraday dispersion function for all lines of sight. 

The use of the window function to alleviate the problem related to incomplete $\lambda^2$ coverage implies that the choice of the window function determines the observation's ability to resolve structures along the Faraday depth axis and the range of Faraday depth's for which the sensitivity is greater than 50\%. If $\delta \lambda^2$ and $\Delta \lambda^2$ are the channel width and the distribution of $\lambda^2$ achieved by the observation, the resolution $(\delta \phi)$ and the maximum Faraday depth for which the sensitivity is greater than 50\% $(||\phi_\mathrm{max}||)$ are 

\begin{equation}
\delta \phi \approx \frac{2 \sqrt{3}}{\Delta \lambda^2}
\end{equation}

\begin{equation}
||\phi_\mathrm{max}|| \approx \frac{\sqrt{3}}{\delta \lambda^2}.
\end{equation}

\subsection{RM synthesis: Computational costs}
From equations \ref{eq:qPhi} and \ref{eq:uPhi}, it is easy to see that $K$, $2\phi$, and $\lambda_i^2-\lambda_0^2$ need to computed only once for the entire dataset and have negligible compute cost. The compute cost to calculate $\widetilde{Q}(\phi)$, $\widetilde{U}(\phi)$, and $\widetilde{F}(\phi)$ for one value of $\phi$ using inputs at one $\lambda^2$ value is 13 FLOP (7 multiplications, 2 trigonometric functions, 2 additions, 1 subtraction and 1 square root operation). 

Thus to calculate $\widetilde{Q}(\phi)$, $\widetilde{U}(\phi)$, and $\widetilde{F}(\phi)$ for one Faraday depth value using $N_\mathrm{chan}$ frequency channels is 
\begin{itemize}
\item $(13+2) \times N_\mathrm{chan}$ FLOP to compute equations \ref{eq:qPhi} -- \ref{eq:pPhi} for a given $\phi$ using all $\lambda_i^2$.
\item $2\left( N_\mathrm{chan} - 1 \right)$ additions for summations over all $i$.
\item 2 multiplications to scale the values by $K$.
\end{itemize}
The above operations need to be repeated $N_\phi$ times to create RM-cubes of $\widetilde{Q}(\phi)$, $\widetilde{U}(\phi)$, and $\widetilde{F}(\phi)$ with $N_\phi$ Faraday depth planes using $N_\mathrm{chan}$ input frequency channels. This amounts to a total computation cost of $15N_\mathrm{chan}N_\phi$ FLOP for a single line of sight. Since each line of sight in the input datacubes can be treated independently, we can scale the computation cost further by the number of spatial pixels in the input datacubes making the final computational cost to be $\sim 15N_\mathrm{chan} N_\phi N_\mathrm{los}$.

For a typical polarimetry data set from the Westerbork Synthesis Radio Telescope, for example from the Westerbork Spitzer Nearby Galaxy Survey (SINGS) survey \citep{braun2007, heald2009b}, which has $512^2$ spatial pixels, 803 frequency channels and 401 Faraday depth planes, the computational cost required to carry out RM synthesis is about 1.27 TFLOP\footnote{1 TFLOP = 1 trillion floating-point operations.}.

To make this estimate, we assumed that each arithmetic and trignometric functions amount to a single floating point operation. In practice, this is however not true. For example, standard libraries evaluate trignometric functions as Taylor series expansions and the number of Taylor terms used in the expansion can be implementation-dependent. Furthermore, the exact number of instructions needed for different mathematical operations available in the standard C libraries can vary depending on the implementation, system hardware and compiler optimization options. Thus the above compute cost should be treated only as a lower limit and not as a precise value.

\section{GPU implementation of RM Synthesis} \label{sec:gpuimplementation}
cuFFS is written in C, and GPU acceleration is achieved using the CUDA Application Programming Interface (API). Our code is publically available through Github\footnote{\url{https://github.com/sarrvesh/cuFFS}}. User input is provided to the program using a configuration file which is parsed using the structured configuration file parser libconfig\footnote{\url{https://github.com/hyperrealm/libconfig}}. Sample input parset file is shown in Figure \ref{fig:inputParset}. 

\begin{figure}[h]
\verbatiminput{parsetFile}
\caption{Sample input parset to \texttt{cuFFS}.}
\label{fig:inputParset}
\end{figure}

Input and output data cubes can either be in FITS \citep{pence1999} or in HDFITS \citep{price2015} file formats. FITS files are read in and written out using the CFITSIO\footnote{\url{https://heasarc.gsfc.nasa.gov/fitsio/}} library while the HDFITS is accessed using the HDF5\footnote{\url{https://support.hdfgroup.org/HDF5/}} library. Justification for supporting data formats in addition to the standard FITS format is provided in section \ref{sec:hdf5}.

A fast CPU implementation of RM synthesis, like the one published in \cite{brentjens2007}, computes the RM cubes by evaluting all Faraday depth values for each input channel. The main advantage of using such a strategy is that it minimizes disk I/O as each input channel is read only once and the output cubes are written to disk only once as long as the output cubes fit in the memory. If all output Faraday depth planes do not fit in memory, the CPU program computes a subset of Faraday depth planes that fits in memory, writes the output planes to disk, and then proceeds to compute the next subset of Faraday depth planes. However, adapting a similar scheme is not very efficient on a GPU as such an implementation would involve numerous time-consuming data transfers between the \textit{host} (CPU) and the \textit{device} (GPU) especially when the output cubes demand larger memory than what is available on the GPU. 

Since data needs to be moved between the CPU and the GPU for processing, computational efficiency can be maximized by minimizing the transfer-to-compute time ratio. This can be achieved by ensuring that a given piece of data is moved between the CPU and the GPU only once. Our GPU implementation ensures this by processing each independent line of sight separately. For a given line of sight, \texttt{cuFFS} carries out the following steps:
\begin{enumerate}
   \item Transfer $\widetilde{Q}(\lambda^2)$ and $\widetilde{U}(\lambda^2)$ to the GPU.
   \item Each thread on the GPU computes $\widetilde{Q}(\phi)$, $\widetilde{U}(\phi)$, and $\widetilde{F}(\phi)$ for a given Faraday depth $\phi$.
   \item Transfer the computed $\widetilde{Q}(\phi)$, $\widetilde{U}(\phi)$, and $\widetilde{F}(\phi)$ to the CPU.
\end{enumerate}
By repeating these steps for each line of sight in the input datacubes, the program constructs the final RM cubes. 

Modern GPUs come with substantial amount of main memory which implies that multiple lines of sight can be processed at the same time. Additionally, transferring a larger chunk of data between host memory and device memory is more efficient than issuing multiple data transfers. As a result, in practice, cuFFS processes multiple lines of sight at once. The program internally estimates the number of lines of sight that can be processed at once based on the amount of memory and the number of processors available on the GPU.

\subsection{Science verification} \label{sec:verification}
To verify the output produced by cuFFS, we compared the output $(\widetilde{F}(\phi))$ produced by cuFFS with a CPU implementation of RM synthesis from \cite{brentjens2007}. The CPU and the GPU codes were used to process a data cube from the Westerbork Synthesis Radio Telescope Spitzer Infrared Nearby Galaxies Survey \citep[WSRT-SINGS; ][]{heald2009b}. We compared the output data cubes $(\widetilde{F}(\phi))$ on a pixel-by-pixel basis and found that the mean of the ratio between the outputs produced by the GPU and the CPU to be $1.0$ with a standard deviation of $1.5 \times 10^{-6}$ implying that the output produced by cuFFS matches the output produced by the CPU code very well. 

\begin{figure*}
\resizebox{\hsize}{!}{\includegraphics[scale=0.95]{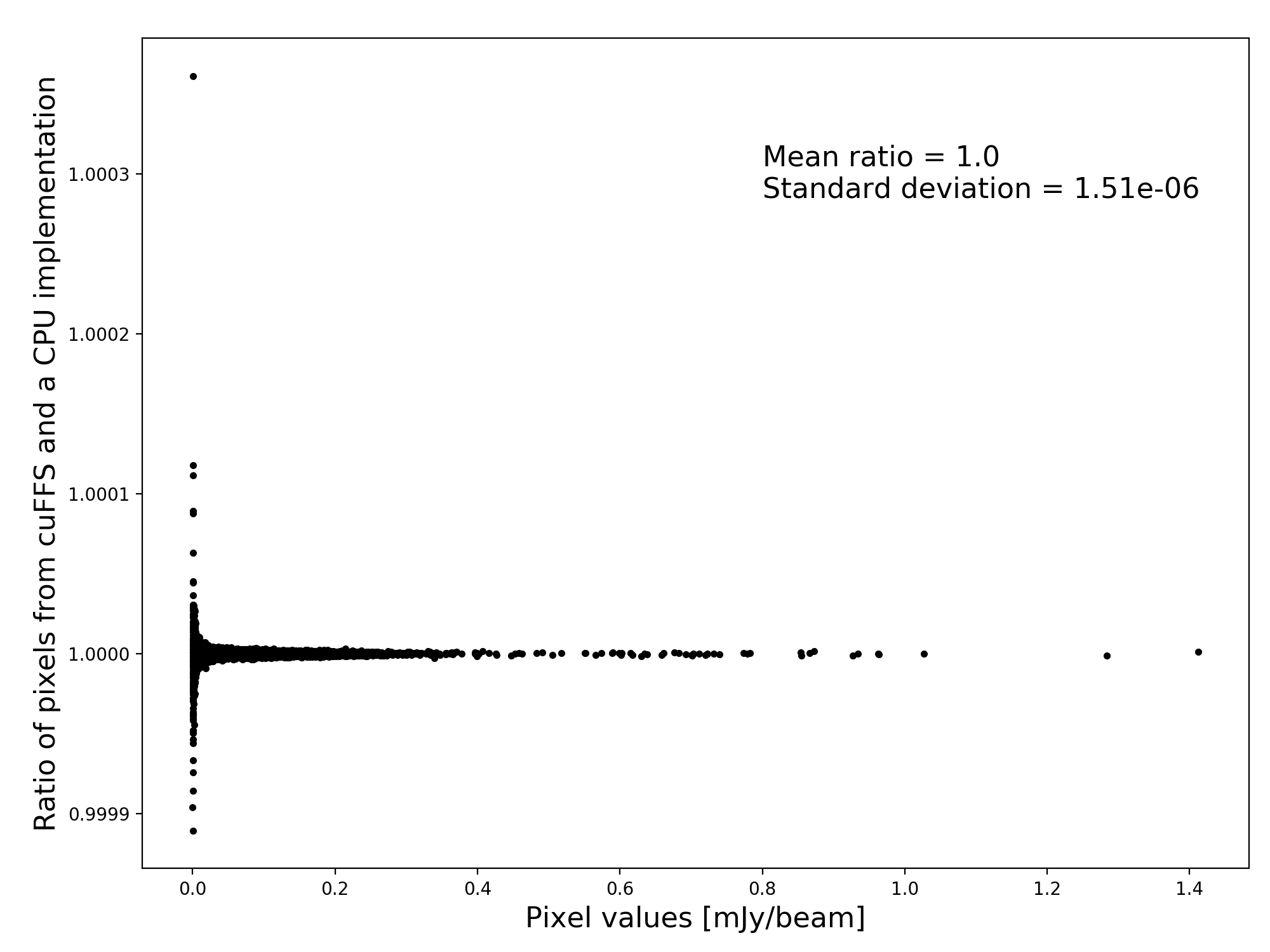}\includegraphics[scale=0.95]{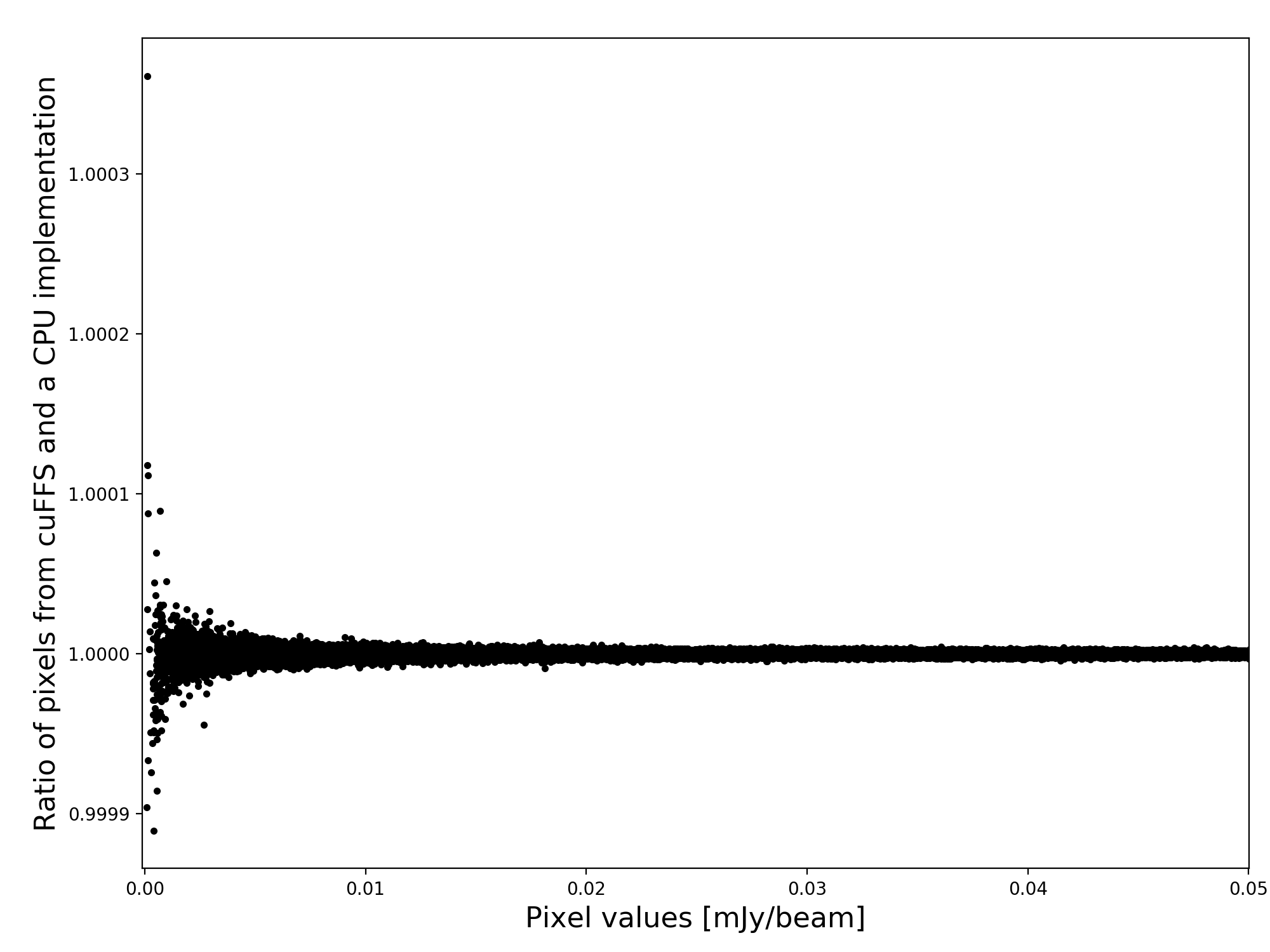}}
\caption{\textbf{Left:} Comparison between the output pixel values produced by cuFFS and a CPU implementation by \cite{brentjens2007}. \textbf{Right:} Zoomed-in version of the figure shown in the left panel to better display the scatter in pixel values computed by the GPU and the CPU codes. The scatter seen for the low output pixel values is probably due to numerical error.}
\label{fig:verify}
\end{figure*}

Figure~\ref{fig:verify} shows the ratio of the pixels values produced by the GPU and the CPU code $(\widetilde{F}_\mathrm{GPU}(\phi) / \widetilde{F}_\mathrm{CPU}(\phi))$ plotted as a function of the output pixel values in $\widetilde{F}_\mathrm{GPU}(\phi)$. As can be seen from Figure~\ref{fig:verify}, the pixel values in the two output cubes match very well.

\subsection{A few comments on file formats} \label{sec:hdf5}
As discussed in the previous section, our code achieves parallelism by exploiting the fact that each line-of-sight and each Faraday depth output for a given line of sight in RM synthesis can be processed independently. However, during the initial development and testing phase, we quickly realized that reading and writing the FITS data cubes along the third data axis (or each line-of-sight) was the major bottleneck in our code. Even for small datacubes like the ones corresponding to WSRT observations, reading and writing datacubes along the third data axis was significantly longer than the amount of time required to carry out the actual RM synthesis computation. In addition to the bottleneck caused by the FITS I/O, it is becoming clear that the FITS data format is not well suited for large multi-dimensional datasets. For an excellent review on the limitations of the FITS data format in the era of big data (radio) astronomy, see \cite{thomas2015}.

One approach to alleviate the slow read/write speed associated with the FITS format would be to rotate (for example using the \texttt{reorder} task in \texttt{miriad} \citep{sault1995}) the input datacubes such that the frequency axis is the fastest-varying dimension. After running cuFFS, the output datacubes, which contain the Faraday depth axis as the fastest-varying dimension, can be derotated back to the original order.

Another approach to reduce the read/write time would be to adopt data formats other than the standard FITS file format. For a recent review on new data formats suited for storing astronomical datacubes, see \cite{shortridge2015}. File formats like the Hierarchical Data Format version 5 (HDF5) are well suited for storing large datacubes and have a significantly lower read/write time compared to the FITS format especially while reading along the slowest-varying axis. Recently, \cite{price2015} proposed a new format called \texttt{HDFITS} which merges the FITS data model with the HDF5 file format. Note that \textit{data model} defines how a dataset is structured and \textit{file format} specifies how the bits are stored on disk. A similar strategy has also been adopted by other data-intensive fields like climatology and oceanography where the \texttt{NetCDF} data format is built on top of the \texttt{HDF5} file format. The main advantage of using \texttt{HDFITS} in our program is that the structure of the dataset stored in the HDF5 file is similar to the standard FITS format with the added advantage of higher read/write speed. 

To achieve higher read/write throughput and to be consistent with existing data format, the current version of \texttt{cuFFS} supports input and output datacubes in both \texttt{FITS} and \texttt{HDFITS} formats. While operating in the \texttt{FITS} mode, \texttt{cuFFS} assumes that the user has rotated the input cubes such that the frequency axis is the fast-varying dimension. Since \texttt{HDFITS} is efficient in reading along the frequency axis (see Table \ref{tab:benchmarks}), no such prior rotation is required while using \texttt{cuFFS} in \texttt{HDFITS} mode. Note that regardless of the input format, the output Faraday depth cubes are ordered such that Faraday depth is the fast-varying dimension.

As mentioned before, numerous data formats have been proposed in the recent years for archiving large datasets. If desired, \texttt{cuFFS} can be modified to support additional data formats like N-dimensional Data Format \citep[NDF;][]{jenness2015a}, Starlink Hierarchical Data System \citep[HDS;][]{jenness2015b} and Advanced Scientific Data Format \citep[ASDF;][]{greenfield2015} with minimal change to the existing code.

\subsection{Benchmarks and performance measurements} \label{sec:benchmarks}
We compared the execution time of our GPU code with a single-threaded and vectorized CPU implementation of RM Synthesis from \cite{brentjens2007} using a combination of both real and mock datasets from current and upcoming telescope facilities. A brief description of each test dataset is presented below:
\begin{itemize}
\item \texttt{WSRT}: As a typical Westerbork polarimetry dataset, we used one of the datacubes produced as part of the Westerbork SINGS survey \cite[WSRT-SINGS; ][]{heald2009b}. After removing frequency channels corrupted by RFI, the final datacube consists of 803 frequency channels. The frequency setup used in the WSRT-SINGS survey results in a Faraday depth resolution of 144~rad~m$^{-2}$. Based on these values, the input and the output datasets have dimensions $512 \times 512 \times 803$ and $512 \times 512 \times 401$ respectively.
\item \texttt{LOFAR\_LR}: Diffuse polarized emission from the Galactic foreground can be detected at 150 MHz in low resolution LOFAR maps with a resolution of a few arcmins \citep{jelic2015,vaneck2017}. The LOFAR low resolution dataset used for this test corresponds to a field around the nearby dwarf galaxy NGC~1569. The input datacubes are at a resolution of $250^{\prime \prime}$. The input Stokes Q and U cubes contain 856 frequency channels which are free of any radio frequency interference. With 48.8 kHz-wide channel maps covering the bandwidth ranging from 120.238 -- 166.087 MHz, the output Faraday cubes will have a Faraday depth resolution of 1.17 rad~m$^{-2}$ and sensitive to a maximum Faraday depth value of about 350~rad~m$^{-2}$.  Thus the dimensions of the input and the output datacubes are $540 \times 540 \times 856$ and $540 \times 540 \times 7000$.
\item \texttt{LOFAR\_HR}: The LOFAR high resolution dataset was produced from the same visibility data that was used to create the \texttt{LOFAR\_LR} datacubes. The only difference between the \texttt{LOFAR\_HR} and \texttt{LOFAR\_LR} is that \texttt{LOFAR\_HR} data cubes are at a higher angular resolution of $26^{\prime \prime}$. The dimensions of the input and the output datacubes are $4000 \times 4000 \times 856$ and $4000 \times 4000 \times 7000$.
\end{itemize}

All CPU and GPU benchmarks were carried out on CSIRO's Bracewell computer cluster located in Canberra, Australia. Each node on the cluster is powered by 28 Intel(R) Xeon(R) CPU E5-2690 v4 \@ 2.60GHz processors with 256 GB RAM and is equipped with four NVidia Tesla P100-SXM2-16GB GPUs. The CPU code was permitted to use up to 200 GB of RAM. The execution times for different datacubes using the CPU and the GPU implementation are listed in Table \ref{tab:benchmarks}. Note that the execution times for the CPU code will increase almost linearly with decreasing RAM size.

From the execution times reported in Table \ref{tab:benchmarks}, it can be seen quite clearly that our GPU implementation in both \texttt{FITS} and \texttt{HDFITS} mode is substantially faster than the execution times achieved using the CPU implementation even though the CPU implementation was allowed to use 15 times more main memory than our GPU program. Execution times reported in Table \ref{tab:benchmarks} while using \texttt{cuFFS} in \texttt{FITS} mode indicate that rotating and derotating the input and output datacubes still dominates the total execution time. If, however, the data rotation is done as part of the imaging pipeline that produces the Stokes Q and U cubes, then the GPU implementation is about two orders of magnitude faster than the CPU implementation for large datasets.

\begin{table*}
\centering
\caption{Comparison of CPU and GPU execution times for different input and output data cubes.}
\label{tab:benchmarks}
\begin{tabular}{@{\extracolsep{2pt}}lllllll@{}}
\hline
Dataset & \multicolumn{3}{c}{Number of} & \multicolumn{3}{c}{Execution time [hh:mm:ss]} \\
\cline{2-4} \cline{5-7}
& Spatial pixels & Channels & $\phi$ planes & CPU$^\mathrm{(a)}$ & GPU$^\mathrm{(a)}$ & GPU$^\mathrm{(b)}$ \\
\hline
WSRT         & $512 \times 512$ & 803 & 401 & 00:02:02 & 00:00:50$^\mathrm{(c)}$+00:00:08$^\mathrm{(d)}$ & 00:00:25\\
LOFAR\_LR    & $540 \times 540$   & 856 & 7000 & 00:39:30 & 00:04:37$^\mathrm{(c)}$+00:00:17$^\mathrm{(d)}$ & 00:04:36 \\
LOFAR\_HR    & $4000 \times 4000$ & 856 & 7000 & 58:01:32 & 12:53:39$^\mathrm{(c)}$+00:13:47$^\mathrm{(d)}$ & 02:28:02 \\
\hline
\end{tabular}
\tablefoot{CPU tests using a single threaded and vectorized code from \cite{brentjens2007} were carried out on a single node with Intel(R) Xeon(R) CPU E5-2690 processor at 2.6 GHz clock with 256 GB RAM. The CPU code was permitted to use a maximum of 200 GB RAM. GPU tests was carried out using an NVIDIA Tesla P100-SXM2-16GB device. (a) FITS file format; (b) HDFITS format (the reported time includes time spent reordering the input datacube and the compute time); (c) Time spent reordering the input datacube to have the frequency axis as the first axis and reordering the output datacubes so that the Faraday depth axis is the third axis; (d) Compute time.}
\end{table*}

\section{Conclusion and future outlook} \label{sec:conclusion}
In this paper, we have presented the GPU-accelerated CUDA program \texttt{cuFFS}, or CUDA-accelerated Fast Faraday Synthesis, to perform a widely used radio astronomy algorithm called rotation measure synthesis. The software package is made publically available through Github. The software package is capable of handling input and output datacubes in either the FITS or the HDFITS format. If needed, additional data formats can also be supported with very little change to the existing code base. In addition to rotation measure synthesis, we plan to implement advanced polarization processing tools like RM-CLEAN \citep{heald2009} in the future version of our software package.

Comparing the execution time of our GPU implementation on NVidia Tesla P100-SXM2-16GB GPU with a single-threaded, vectorized CPU implementation, we observe gains in execution times by at least a factor of 15 for smaller datasets from telescopes like WSRT and by at least by two orders of magnitude for larger datasets from telescopes like LOFAR, depending on the input and output data format. Further increase in speed can be achieved by using Fast Fourier Transforms (FFT) instead of Discrete Fourier Transforms (DFT) in equations \ref{eq:qPhi} and \ref{eq:uPhi}. CUDA accelerated Fast Fourier Transform can be accelerated using standard libraries like \texttt{cuFFT}\footnote{\url{https://developer.nvidia.com/cufft}}. We aim to implement these advanced features in future versions of \texttt{cuFFS}.

Finally, as discussed in section \ref{sec:benchmarks} and Table \ref{tab:benchmarks}, the main bottleneck in our implementation of RM synthesis is the FITS file access using CFITSIO. Even for a small WSRT dataset, rotating the cubes to speed up I/O can be an order of magnitude longer than the actual compute time. In light of the large data cubes that will be produced by the current and upcoming radio telescopes, this I/O overload is certainly a cause for concern. To minimize the amount of time spent rearranging data, we make the following two recommendations to teams carrying out large polarization surveys:
\begin{itemize}
   \item Adopt data formats like HDFITS which benefit from combining the FITS data model with the HDF5 file format. Accessing an array stored in HDFITS format along the slowest-varying dimension can be two orders of magnitude faster than FITS format.
   \item If surveys prefer to use the standard FITS format to store their data products, they should at least consider archiving the data products not intended for visualization (like $Q(\lambda)$, $U(\lambda)$, $Q(\phi)$, and $U(\phi)$ cubes) with frequency and Faraday depth as the first/fastest-varying axis.
\end{itemize}

\section*{Acknowledgements}
\noindent
We thank the anonymous referee for their helpful comments that improved the manuscript. SSS acknowledges financial support from the Leids Kerkhoven-Bosscha Fonds (LKBF) for multiple visits to CSIRO in Perth, Australia where most of this work was carried out. SSS thanks Christopher Harris at the Pawsey Supercomputing Centre for useful discussions on CUDA programming. JMvdH acknowledges support from the European Research Council under the European Union's Seventh Frame-work Programme (FP/2007-2013)/ERC Grant Agreement no. 291531. We thank the IMT Scientific Computing group at CSIRO for their expertise and support of our work on the CSIRO Bracewell GPU cluster.

\bibliographystyle{model2-names}
\bibliography{rmsynthesis}

\begin{thebibliography}{42}
\expandafter\ifx\csname natexlab\endcsname\relax\def\natexlab#1{#1}\fi
\expandafter\ifx\csname url\endcsname\relax
  \def\url#1{\texttt{#1}}\fi
\expandafter\ifx\csname urlprefix\endcsname\relax\def\urlprefix{URL }\fi

\bibitem[{{Bassa} et~al.(2017){Bassa}, {Pleunis}, and {Hessels}}]{bassa2017}
{Bassa}, C.~G., {Pleunis}, Z., {Hessels}, J.~W.~T., Jan. 2017. {Enabling pulsar
  and fast transient searches using coherent dedispersion}. Astronomy and
  Computing 18, 40--46.

\bibitem[{{Beck}(2016)}]{beck2016}
{Beck}, R., Dec. 2016. {Magnetic fields in spiral galaxies}. \aapr 24, 4.

\bibitem[{Bouchez et~al.(2012)Bouchez, Actonb, Biasic, Conand, Espelanda,
  Espositoe, Filgueiraa, Gallienif, McLeodg, Pinnae, et~al.}]{bouchez2012}
Bouchez, A.~H., Actonb, D.~S., Biasic, R., Conand, R., Espelanda, B.,
  Espositoe, S., Filgueiraa, J., Gallienif, D., McLeodg, B.~A., Pinnae, E.,
  et~al., 2012. The giant magellan telescope adaptive optics program. In: Proc.
  of SPIE Vol. Vol. 9148. pp. 91480W--1.

\bibitem[{{Braun} et~al.(2007){Braun}, {Oosterloo}, {Morganti}, {Klein}, and
  {Beck}}]{braun2007}
{Braun}, R., {Oosterloo}, T.~A., {Morganti}, R., {Klein}, U., {Beck}, R., Jan.
  2007. {The Westerbork SINGS survey. I. Overview and image atlas}. \aap 461,
  455--470.

\bibitem[{{Brentjens}(2007)}]{brentjens2007}
{Brentjens}, M.~A., 2007. {Radio polarimetry in 2.5D}. Ph.D. thesis, University
  of Groningen.

\bibitem[{{Brentjens} and {de Bruyn}(2005)}]{brentjens2005}
{Brentjens}, M.~A., {de Bruyn}, A.~G., Oct. 2005. {Faraday rotation measure
  synthesis}. \aap 441, 1217--1228.

\bibitem[{{Burn}(1966)}]{burn1966}
{Burn}, B.~J., 1966. {On the depolarization of discrete radio sources by
  Faraday dispersion}. \mnras 133, 67.

\bibitem[{{Chennamangalam} et~al.(2014){Chennamangalam}, {Scott}, {Jones},
  {Chen}, {Ford}, {Kepley}, {Lorimer}, {Nie}, {Prestage}, {Roshi}, {Wagner},
  and {Werthimer}}]{chennamangalam2014}
{Chennamangalam}, J., {Scott}, S., {Jones}, G., {Chen}, H., {Ford}, J.,
  {Kepley}, A., {Lorimer}, D.~R., {Nie}, J., {Prestage}, R., {Roshi}, D.~A.,
  {Wagner}, M., {Werthimer}, D., Dec. 2014. {A GPU-Based Wide-Band Radio
  Spectrometer}. \pasa 31, e048.

\bibitem[{{Elsen} et~al.(2007){Elsen}, {Vishal}, {Houston}, {Pande},
  {Hanrahan}, and {Darve}}]{elsen2007}
{Elsen}, E., {Vishal}, V., {Houston}, M., {Pande}, V., {Hanrahan}, P., {Darve},
  E., Jun. 2007. {N-Body Simulations on GPUs}. ArXiv e-prints.

\bibitem[{{Fletcher} and {Klein}(2015)}]{fletcher2015}
{Fletcher}, A., {Klein}, U., 2015. {Galactic and Intergalactic Magnetic
  Fields}. Heidelberg: Springer.

\bibitem[{{Fluke}(2012)}]{fluke2012}
{Fluke}, C.~J., Sep. 2012. {Accelerating the Rate of Astronomical Discovery
  with GPU-Powered Clusters}. In: {Ballester}, P., {Egret}, D., {Lorente},
  N.~P.~F. (Eds.), Astronomical Data Analysis Software and Systems XXI. Vol.
  461 of Astronomical Society of the Pacific Conference Series. p.~3.

\bibitem[{{Fluke} et~al.(2011){Fluke}, {Barnes}, {Barsdell}, and
  {Hassan}}]{fluke2011}
{Fluke}, C.~J., {Barnes}, D.~G., {Barsdell}, B.~R., {Hassan}, A.~H., Jan. 2011.
  {Astrophysical Supercomputing with GPUs: Critical Decisions for Early
  Adopters}. \pasa 28, 15--27.

\bibitem[{{Greenfield} et~al.(2015){Greenfield}, {Droettboom}, and
  {Bray}}]{greenfield2015}
{Greenfield}, P., {Droettboom}, M., {Bray}, E., Sep. 2015. {ASDF: A new data
  format for astronomy}. Astronomy and Computing 12, 240--251.

\bibitem[{{Hamaker} et~al.(1996){Hamaker}, {Bregman}, and
  {Sault}}]{hamaker1996}
{Hamaker}, J.~P., {Bregman}, J.~D., {Sault}, R.~J., May 1996. {Understanding
  radio polarimetry. I. Mathematical foundations.} \aaps 117, 137--147.

\bibitem[{{Harris} et~al.(2008){Harris}, {Haines}, and
  {Staveley-Smith}}]{harris2008}
{Harris}, C., {Haines}, K., {Staveley-Smith}, L., Oct. 2008. {GPU accelerated
  radio astronomy signal convolution}. Experimental Astronomy 22, 129--141.

\bibitem[{{Heald}(2009)}]{heald2009}
{Heald}, G., Apr. 2009. {The Faraday rotation measure synthesis technique}. In:
  {Strassmeier}, K.~G., {Kosovichev}, A.~G., {Beckman}, J.~E. (Eds.), IAU
  Symposium. Vol. 259 of IAU Symposium. pp. 591--602.

\bibitem[{{Heald} et~al.(2009){Heald}, {Braun}, and {Edmonds}}]{heald2009b}
{Heald}, G., {Braun}, R., {Edmonds}, R., Aug. 2009. {The Westerbork SINGS
  survey. II Polarization, Faraday rotation, and magnetic fields}. \aap 503,
  409--435.

\bibitem[{{Jeli{\'c}} et~al.(2015){Jeli{\'c}}, {de Bruyn}, {Pandey}, {Mevius},
  {Haverkorn}, {Brentjens}, {Koopmans}, {Zaroubi}, {Abdalla}, {Asad}, {Bus},
  {Chapman}, {Ciardi}, {Fernandez}, {Ghosh}, {Harker}, {Iliev}, {Jensen},
  {Kazemi}, {Mellema}, {Offringa}, {Patil}, {Vedantham}, and
  {Yatawatta}}]{jelic2015}
{Jeli{\'c}}, V., {de Bruyn}, A.~G., {Pandey}, V.~N., {Mevius}, M., {Haverkorn},
  M., {Brentjens}, M.~A., {Koopmans}, L.~V.~E., {Zaroubi}, S., {Abdalla},
  F.~B., {Asad}, K.~M.~B., {Bus}, S., {Chapman}, E., {Ciardi}, B., {Fernandez},
  E.~R., {Ghosh}, A., {Harker}, G., {Iliev}, I.~T., {Jensen}, H., {Kazemi}, S.,
  {Mellema}, G., {Offringa}, A.~R., {Patil}, A.~H., {Vedantham}, H.~K.,
  {Yatawatta}, S., Nov. 2015. {Linear polarization structures in LOFAR
  observations of the interstellar medium in the 3C 196 field}. \aap 583, A137.

\bibitem[{{Jenness}(2015)}]{jenness2015b}
{Jenness}, T., Sep. 2015. {Reimplementing the Hierarchical Data System using
  HDF5}. Astronomy and Computing 12, 221--228.

\bibitem[{{Jenness} et~al.(2015){Jenness}, {Berry}, {Currie}, {Draper},
  {Economou}, {Gray}, {McIlwrath}, {Shortridge}, {Taylor}, {Wallace}, and
  {Warren-Smith}}]{jenness2015a}
{Jenness}, T., {Berry}, D.~S., {Currie}, M.~J., {Draper}, P.~W., {Economou},
  F., {Gray}, N., {McIlwrath}, B., {Shortridge}, K., {Taylor}, M.~B.,
  {Wallace}, P.~T., {Warren-Smith}, R.~F., Sep. 2015. {Learning from 25 years
  of the extensible N-Dimensional Data Format}. Astronomy and Computing 12,
  146--161.

\bibitem[{{Magro} et~al.(2011){Magro}, {Karastergiou}, {Salvini}, {Mort},
  {Dulwich}, and {Zarb Adami}}]{magro2011}
{Magro}, A., {Karastergiou}, A., {Salvini}, S., {Mort}, B., {Dulwich}, F.,
  {Zarb Adami}, K., Nov. 2011. {Real-time, fast radio transient searches with
  GPU de-dispersion}. \mnras 417, 2642--2650.

\bibitem[{Michalakes and Vachharajani(2008)}]{michalakes2008}
Michalakes, J., Vachharajani, M., 2008. Gpu acceleration of numerical weather
  prediction. Parallel Processing Letters 18~(04), 531--548.

\bibitem[{Nvidia(2017)}]{cuda}
Nvidia, 2017. {CUDA Programming Guide 8.0}.

\bibitem[{{Ord} et~al.(2009){Ord}, {Greenhill}, {Wayth}, {Mitchell}, {Dale},
  {Pfister}, and {Edgar}}]{ord2009}
{Ord}, S., {Greenhill}, L., {Wayth}, R., {Mitchell}, D., {Dale}, K., {Pfister},
  H., {Edgar}, R., Sep. 2009. {Graphics Processing Units for Data Processing in
  the Murchison Wide-field Array}. In: {Bohlender}, D.~A., {Durand}, D.,
  {Dowler}, P. (Eds.), Astronomical Data Analysis Software and Systems XVIII.
  Vol. 411 of Astronomical Society of the Pacific Conference Series. p. 127.

\bibitem[{Owens et~al.(2008)Owens, Houston, Luebke, Green, Stone, and
  Phillips}]{owens2008}
Owens, J.~D., Houston, M., Luebke, D., Green, S., Stone, J.~E., Phillips,
  J.~C., 2008. Gpu computing. Proceedings of the IEEE 96~(5), 879--899.

\bibitem[{Owens et~al.(2007)Owens, Luebke, Govindaraju, Harris, Krüger,
  Lefohn, and Purcell}]{owens2007}
Owens, J.~D., Luebke, D., Govindaraju, N., Harris, M., Krüger, J., Lefohn,
  A.~E., Purcell, T.~J., 2007. A survey of general-purpose computation on
  graphics hardware. Computer Graphics Forum 26~(1), 80--113.

\bibitem[{Patterson and Hennessy(2014)}]{patterson2014}
Patterson, D., Hennessy, J., 2014. Computer Organization and Design, Enhanced:
  The Hardware/Software Interface. The Morgan Kaufmann Series in Computer
  Architecture and Design. Elsevier Science.
\newline\urlprefix\url{https://books.google.nl/books?id=G7IMAwAAQBAJ}

\bibitem[{{Pence}(1999)}]{pence1999}
{Pence}, W., 1999. {CFITSIO, v2.0: A New Full-Featured Data Interface}. In:
  {Mehringer}, D.~M., {Plante}, R.~L., {Roberts}, D.~A. (Eds.), Astronomical
  Data Analysis Software and Systems VIII. Vol. 172 of Astronomical Society of
  the Pacific Conference Series. p. 487.

\bibitem[{{Perkins} et~al.(2015){Perkins}, {Marais}, {Zwart}, {Natarajan},
  {Tasse}, and {Smirnov}}]{perkins2015}
{Perkins}, S.~J., {Marais}, P.~C., {Zwart}, J.~T.~L., {Natarajan}, I., {Tasse},
  C., {Smirnov}, O., Sep. 2015. {Montblanc$^{1}$: GPU accelerated radio
  interferometer measurement equations in support of Bayesian inference for
  radio observations}. Astronomy and Computing 12, 73--85.

\bibitem[{{Portegies Zwart} et~al.(2007){Portegies Zwart}, {Belleman}, and
  {Geldof}}]{zwart2007}
{Portegies Zwart}, S.~F., {Belleman}, R.~G., {Geldof}, P.~M., Nov. 2007.
  {High-performance direct gravitational N-body simulations on graphics
  processing units}. \na 12, 641--650.

\bibitem[{{Price} et~al.(2015){Price}, {Barsdell}, and {Greenhill}}]{price2015}
{Price}, D.~C., {Barsdell}, B.~R., {Greenhill}, L.~J., Sep. 2015. {HDFITS:
  Porting the FITS data model to HDF5}. Astronomy and Computing 12, 212--220.

\bibitem[{{Price} et~al.(2016){Price}, {Staveley-Smith}, {Bailes}, {Carretti},
  {Jameson}, {Jones}, {van Straten}, and {Schediwy}}]{price2016}
{Price}, D.~C., {Staveley-Smith}, L., {Bailes}, M., {Carretti}, E., {Jameson},
  A., {Jones}, M.~E., {van Straten}, W., {Schediwy}, S.~W., Dec. 2016. {HIPSR:
  A Digital Signal Processor for the Parkes 21-cm Multibeam Receiver}. Journal
  of Astronomical Instrumentation 5, 1641007.

\bibitem[{{Sault} et~al.(1995){Sault}, {Teuben}, and {Wright}}]{sault1995}
{Sault}, R.~J., {Teuben}, P.~J., {Wright}, M.~C.~H., 1995. {A Retrospective
  View of MIRIAD}. In: {Shaw}, R.~A., {Payne}, H.~E., {Hayes}, J.~J.~E. (Eds.),
  Astronomical Data Analysis Software and Systems IV. Vol.~77 of Astronomical
  Society of the Pacific Conference Series. p. 433.

\bibitem[{{Schaaf} and {Overeem}(2004)}]{schaaf2004}
{Schaaf}, K., {Overeem}, R., Jun. 2004. {Cots Correlator Platform}.
  Experimental Astronomy 17, 287--297.

\bibitem[{Schiwietz et~al.(2006)Schiwietz, Chang, Speier, and
  Westermann}]{schiwietz2006}
Schiwietz, T., Chang, T.-c., Speier, P., Westermann, R., 2006. Mr image
  reconstruction using the gpu. In: Proc. of SPIE. Vol. 6142. p. 61423T.

\bibitem[{Schomberg and Timmer(1995)}]{schomberg1995}
Schomberg, H., Timmer, J., 1995. The gridding method for image reconstruction
  by fourier transformation. IEEE transactions on medical imaging 14~(3),
  596--607.

\bibitem[{{Shortridge}(2015)}]{shortridge2015}
{Shortridge}, K., Sep. 2015. {A Glimpse at Different File Formats}. In:
  {Taylor}, A.~R., {Rosolowsky}, E. (Eds.), Astronomical Data Analysis Software
  an Systems XXIV (ADASS XXIV). Vol. 495 of Astronomical Society of the Pacific
  Conference Series. p. 527.

\bibitem[{{Thomas} et~al.(2015){Thomas}, {Jenness}, {Economou}, {Greenfield},
  {Hirst}, {Berry}, {Bray}, {Gray}, {Muna}, {Turner}, {de Val-Borro},
  {Santander-Vela}, {Shupe}, {Good}, {Berriman}, {Kitaeff}, {Fay}, {Laurino},
  {Alexov}, {Landry}, {Masters}, {Brazier}, {Schaaf}, {Edwards}, {Redman},
  {Marsh}, {Streicher}, {Norris}, {Pascual}, {Davie}, {Droettboom},
  {Robitaille}, {Campana}, {Hagen}, {Hartogh}, {Klaes}, {Craig}, and
  {Homeier}}]{thomas2015}
{Thomas}, B., {Jenness}, T., {Economou}, F., {Greenfield}, P., {Hirst}, P.,
  {Berry}, D.~S., {Bray}, E., {Gray}, N., {Muna}, D., {Turner}, J., {de
  Val-Borro}, M., {Santander-Vela}, J., {Shupe}, D., {Good}, J., {Berriman},
  G.~B., {Kitaeff}, S., {Fay}, J., {Laurino}, O., {Alexov}, A., {Landry}, W.,
  {Masters}, J., {Brazier}, A., {Schaaf}, R., {Edwards}, K., {Redman}, R.~O.,
  {Marsh}, T.~R., {Streicher}, O., {Norris}, P., {Pascual}, S., {Davie}, M.,
  {Droettboom}, M., {Robitaille}, T., {Campana}, R., {Hagen}, A., {Hartogh},
  P., {Klaes}, D., {Craig}, M.~W., {Homeier}, D., Sep. 2015. {Learning from
  FITS: Limitations in use in modern astronomical research}. Astronomy and
  Computing 12, 133--145.

\bibitem[{Trancoso and Charalambous(2005)}]{trancoso2005}
Trancoso, P., Charalambous, M., 2005. Exploring graphics processor performance
  for general purpose applications. In: Digital System Design, 2005.
  Proceedings. 8th Euromicro Conference on. IEEE, pp. 306--313.

\bibitem[{{Vallee}(1980)}]{vallee1980}
{Vallee}, J.~P., Jun. 1980. {The rotation measures of radio sources and their
  data processing}. \aap 86, 251--253.

\bibitem[{{Van Eck} et~al.(2017){Van Eck}, {Haverkorn}, {Alves}, {Beck}, {de
  Bruyn}, {En{\ss}lin}, {Farnes}, {Ferri{\`e}re}, {Heald}, {Horellou},
  {Horneffer}, {Iacobelli}, {Jeli{\'c}}, {Mart{\'{\i}}-Vidal}, {Mulcahy},
  {Reich}, {R{\"o}ttgering}, {Scaife}, {Schnitzeler}, {Sobey}, and
  {Sridhar}}]{vaneck2017}
{Van Eck}, C.~L., {Haverkorn}, M., {Alves}, M.~I.~R., {Beck}, R., {de Bruyn},
  A.~G., {En{\ss}lin}, T., {Farnes}, J.~S., {Ferri{\`e}re}, K., {Heald}, G.,
  {Horellou}, C., {Horneffer}, A., {Iacobelli}, M., {Jeli{\'c}}, V.,
  {Mart{\'{\i}}-Vidal}, I., {Mulcahy}, D.~D., {Reich}, W., {R{\"o}ttgering},
  H.~J.~A., {Scaife}, A.~M.~M., {Schnitzeler}, D.~H.~F.~M., {Sobey}, C.,
  {Sridhar}, S.~S., Jan. 2017. {Faraday tomography of the local interstellar
  medium with LOFAR: Galactic foregrounds towards IC 342}. \aap 597, A98.

\bibitem[{{Zink}(2011)}]{zink2011}
{Zink}, B., Feb. 2011. {HORIZON: Accelerated General Relativistic
  Magnetohydrodynamics}. ArXiv e-prints.

\end{thebibliography}

\end{document}